\documentclass{iau307}
\usepackage{graphicx}
\usepackage{natbib}
\usepackage{url}
\usepackage{dtklogos}
\bibpunct{(}{)}{;}{a}{}{,}

\title[Internal mixing of main sequence B stars] 
{Magnetic fields and internal mixing \\of main sequence B stars}

\author[G.A. Wade et al.]   
{G.A. Wade$^1$, C.P. Folsom$^2$, J. Grunhut$^3$, J.D. Landstreet$^{4,5}$, V. Petit$^6$}

\affiliation{$^1$RMC, Canada, $^2$IRAP, France, $^3$ESO, Germany, $^4$UWO, Canada, $^5$Armagh, U.K., $^6$U. Delaware, USA}

\pubyear{2014}
\volume{307} 
\pagerange{}
\jname{New windows on massive stars: asteroseismology, interferometry, and spectropolarimetry}
\editors{G. Meynet, C. Georgy, J.H. Groh \& Ph. Stee, eds.}

\begin{document}

\maketitle

\begin{abstract}
We have obtained high-quality magnetic field measurements of 19 sharp-lined B-type stars with precisely-measured N/C abundance ratios \citep{2012A&A...539A.143N}.  Our primary goal is to test the idea \citep{2011A&A...525L..11M} that a magnetic field may explain extra drag (through the wind) on the surface rotation, thus producing more internal shear and mixing, and thus could provide an explanation for the appearance of slowly rotating N-rich main sequence B stars. 
\keywords{stars: abundances, stars: early-type, stars: evolution, stars: magnetic fields}
\end{abstract}

\firstsection 

\section{Introduction and motivation}

Observations of chemically-enriched main sequence stars \citep[e.g.][]{1992ApJ...387..673G} have led to the idea that rotationally induced mixing may bring fusion products from the cores of massive stars to their surfaces during phases of core hydrogen burning. Measurements of surface chemical abundances - in particular those of light elements - of early-type main sequence stars can therefore be leveraged to place constraints on the associated circulation currents. A surprising result of these studies is the identification of a significant population of stars with enhanced nitrogen abundances with apparently slow rotation \citep{2008ApJ...676L..29H,2006A&A...457..651M}. The origin of the nitrogen enhancement in these stars is not understood, and may be related to various physical processes, e.g. true slow rotation, pulsation \citep{2014ApJ...781...88A}, binarity \citep{2008IAUS..252..467L} or magnetic fields \citep{2008A&A...481..453M}. 

\begin{figure}[b]
\begin{center}
\includegraphics[width=0.4\textwidth,angle=-90]{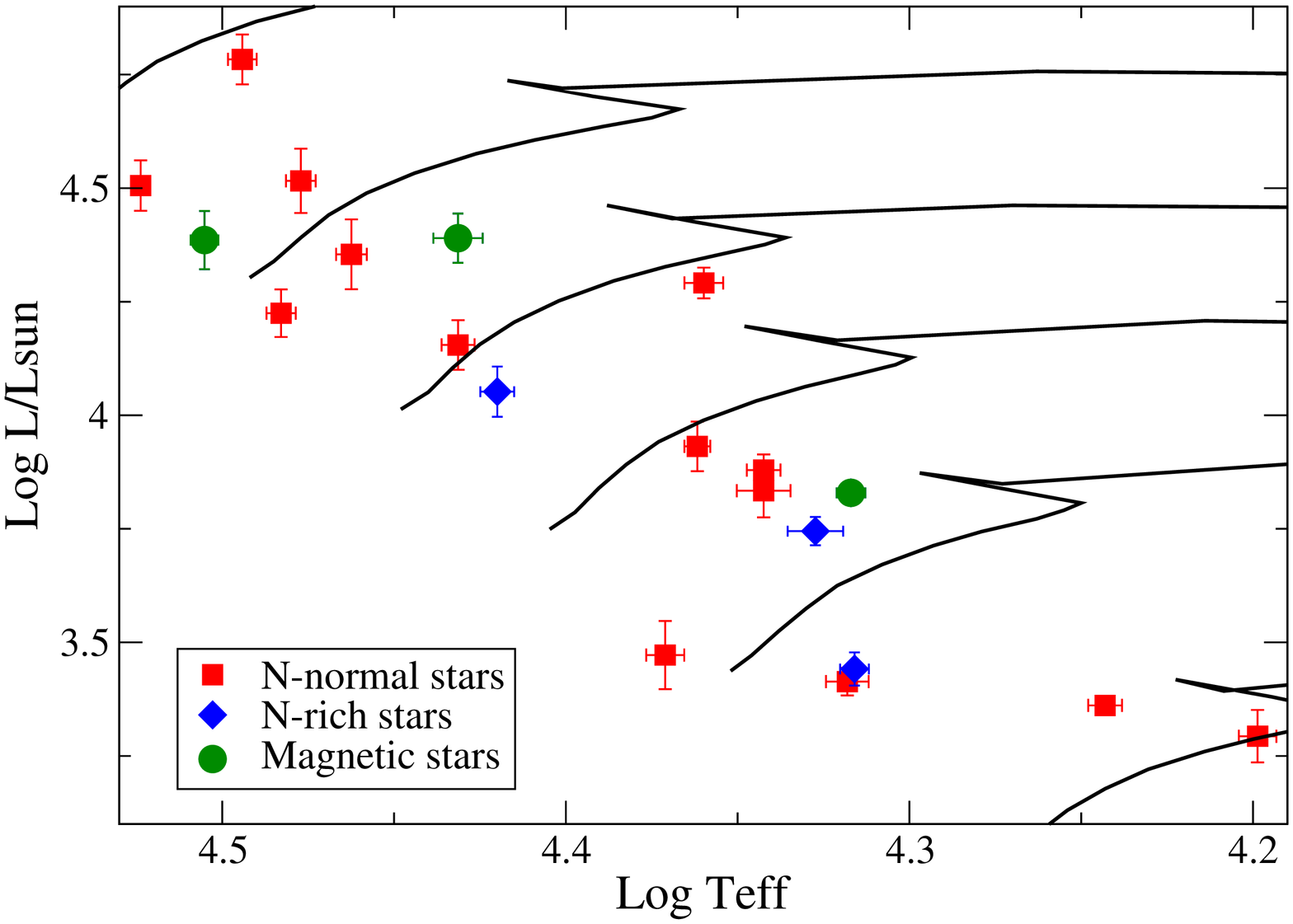}\includegraphics[width=0.4\textwidth,angle=-90]{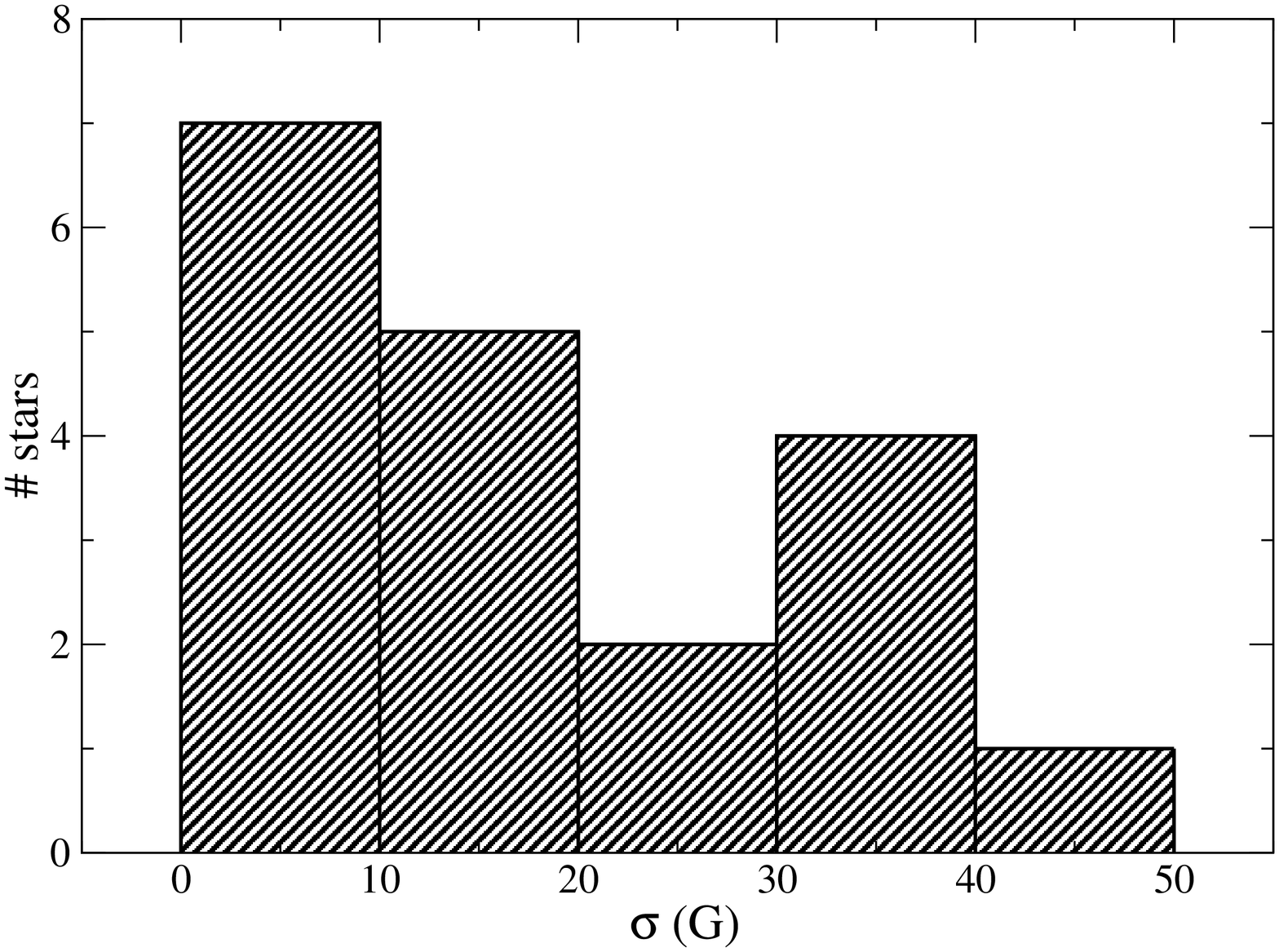}
\caption{{\em Left panel:}\ HR diagram of the sample stars. Evolutionary tracks \citep{1992A&AS...96..269S} span 6-25~$M_\odot$. {\em Right panel:}\ Histogram of longitudinal magnetic field measurement standard errors.}
\label{fig1}
\end{center}
\end{figure}

\section{The sample}

Our sample consists of 19 of the the 20 nearby main sequence B stars for which \citet{2012A&A...539A.143N} determined precise, homogeneous abundances of CNO elements. All of the sample stars are sharp-lined, and 3 stars ($\zeta$~Cas, $\tau$~Sco, and $\beta$~Cep) are known magnetic stars. The HR diagram of the sample is shown in Fig. 1. We identify 6 N-rich stars, those with N/C differing by $>2\sigma$ from the solar value of $-0.60\pm 0.07$ (Asplund et al. 2009): the 3 magnetic stars, as well as HD 61068 (N/C$=-0.27$, $\sigma_B=9$~G), HD 16582 (N/C=$+0.02$, $\sigma_B=3$~G) and HD 35708 (N/C=$-0.08$, $\sigma_B=8$~G).

\section{Magnetic diagnosis}

Magnetic fields were diagnosed for 19 stars using high resolution Stokes $V$ spectropolarimetry obtained using the ESPaDOnS, Narval and Harpspol spectropolarimeters. Some observations were obtained as part of the MiMeS Large Programs, while others were obtained from a dedicated PI program. Least-Squares Deconvolution (LSD) using tailored line masks was used to obtain high SNR mean Stokes $I$, $V$ and diagnostic null line profiles for each star. Typically, 1-3 observations per target were obtained. The presence of a magnetic field was evaluated using the $\chi^2$ detection criterion of \citet{1997MNRAS.291..658D}, as well as via measurement of the longitudinal magnetic field. No new magnetic stars are detected, with a median longitudinal field error bar of 13~G (see histogram in Fig. 1). We have also inferred upper limits on the surface dipole fields of the non-magnetic stars using the method of \citet{2012MNRAS.420..773P}. All odds ratios favour the non-magnetic model, and all probability density functions characterizing the dipole field strength are consistent with zero within 68\% confidence.

\section{Results, conclusions and future work}

We find that all magnetic stars in the sample show significant N enrichment. We therefore conclude that the proposed mechanism by which magnetic braking increases mixing from the deep interior \citep{2011A&A...525L..11M} does seem to be supported by our analysis. However, \citet{2011IAUS..272...97M} reported normal N/C ratios for two magnetic early-type stars (NGC 2244-201, HD 57682), albeit with somewhat worse precision. Taking this at face value, the results appear presently to be ambiguous. Moreover, a small number of confidently non-magnetic stars in our sample are also observed to be significantly N-rich. In particular, the non-magnetic star HD 16582 (B2IV) shows the largest N/C enrichment of the sample. To explain the non-magnetic N-rich stars, additional mechanisms must be at play. Considering the small size of the current sample, an obvious extension will be to obtain similarly precise N/C ratios and magnetic data for a much larger sample of magnetic and non-magnetic early B/late O stars.

\bibliographystyle{iau307}
\bibliography{IAUS307_Wade_Nabuns}

\begin{thebibliography}{}

\bibitem[\protect\astroncite{{Aerts} et~al.}{2014}]{2014ApJ...781...88A}
{Aerts}, C., {Molenberghs}, G., {Kenward}, M.~G., \& {Neiner}, C. 2014,
\newblock {\em \apj} 781, 88

\bibitem[\protect\astroncite{{Donati} et~al.}{1997}]{1997MNRAS.291..658D}
{Donati}, J.-F., {Semel}, M., {Carter}, B.~D., {Rees}, D.~E., \& {Collier
  Cameron}, A. 1997,
\newblock {\em \mnras} 291, 658

\bibitem[\protect\astroncite{{Gies} \& {Lambert}}{1992}]{1992ApJ...387..673G}
{Gies}, D.~R. \& {Lambert}, D.~L. 1992,
\newblock {\em \apj} 387, 673

\bibitem[\protect\astroncite{{Hunter} et~al.}{2008}]{2008ApJ...676L..29H}
{Hunter}, I., {Brott}, I., {Lennon}, D.~J., {et~al.} 2008,
\newblock {\em \apjl} 676, L29

\bibitem[\protect\astroncite{{Langer}}{2008}]{2008IAUS..252..467L}
{Langer}, N. 2008,
\newblock in L. {Deng} \& K.~L. {Chan} (eds.), {\em IAU Symposium}, Vol. 252 of
  {\em IAU Symposium}, pp 467--473

\bibitem[\protect\astroncite{{Meynet} et~al.}{2011}]{2011A&A...525L..11M}
{Meynet}, G., {Eggenberger}, P., \& {Maeder}, A. 2011,
\newblock {\em \aap} 525, L11

\bibitem[\protect\astroncite{{Morel}}{2011}]{2011IAUS..272...97M}
{Morel}, T. 2011,
\newblock in C. {Neiner}, G. {Wade}, G. {Meynet}, \& G. {Peters} (eds.), {\em
  IAU Symposium}, Vol. 272 of {\em IAU Symposium}, pp 97--98

\bibitem[\protect\astroncite{{Morel} et~al.}{2006}]{2006A&A...457..651M}
{Morel}, T., {Butler}, K., {Aerts}, C., {Neiner}, C., \& {Briquet}, M. 2006,
\newblock {\em \aap} 457, 651

\bibitem[\protect\astroncite{{Morel} et~al.}{2008}]{2008A&A...481..453M}
{Morel}, T., {Hubrig}, S., \& {Briquet}, M. 2008,
\newblock {\em \aap} 481, 453

\bibitem[\protect\astroncite{{Nieva} \&
  {Przybilla}}{2012}]{2012A&A...539A.143N}
{Nieva}, M.-F. \& {Przybilla}, N. 2012,
\newblock {\em \aap} 539, A143

\bibitem[\protect\astroncite{{Petit} \& {Wade}}{2012}]{2012MNRAS.420..773P}
{Petit}, V. \& {Wade}, G.~A. 2012,
\newblock {\em \mnras} 420, 773

\bibitem[\protect\astroncite{{Schaller} et~al.}{1992}]{1992A&AS...96..269S}
{Schaller}, G., {Schaerer}, D., {Meynet}, G., \& {Maeder}, A. 1992,
\newblock {\em \aaps} 96, 269

\end{thebibliography}

%
%
%
%
%

\end{document}